\documentclass[twocolumn,10pt]{article}

\usepackage[utf8]{inputenc}
\usepackage[T1]{fontenc}
\usepackage{graphicx}
\usepackage{amsmath, amssymb, amsfonts}
\usepackage{color}
\usepackage{url}
\usepackage[margin=1.8cm]{geometry} 

\newcommand{\keywords}[1]{\par\addvspace\baselineskip\noindent\textbf{Keywords: }#1}

\title{\textbf{A Yang-Mills Type Gauge Theory of Gravity and the Dark Matter Problem}}

\author{
    Yi Yang$^{1,2,}$\thanks{e-mail: yiyang429@as.edu.tw} \and 
    Wai Bong Yeung$^{1}$
}

\date{\small $^{1}$Institute of Physics, Academia Sinica, Taipei, 11529, Taiwan, ROC \\
      $^{2}$Department of Physics, National Cheng Kung University, Tainan, 701, Taiwan, ROC \\
      \vspace{1em}
      \today}

\begin{document}

\maketitle

\begin{abstract}
A Yang-Mills type gauge theory of gravity is shown to have a structure richer than that of Einstein's General Theory of Relativity. By elevating the full connections to independent dynamical gauge fields, the theory admits non-trivial vacuum configurations driven by gauge field self-interactions. We demonstrate that this non-linear geometric structure naturally yields an effective spacetime that explains the form of the galactic rotation curves and the anomalous amount of intergalactic gravitational lensing, completely circumventing the need for exotic dark matter particles.
\end{abstract}

\section{Introduction}
In the past decades, many new discoveries in astronomy that might have something to do with gravity are coming in. For example, stellar objects at the spiral arms of galaxies are rotating at faster speeds than that can be explained by the Keplerian motions. To overcome this difficulty, people assume that some extra matter, not visible to us, is giving an extra pull on these stellar objects. Extra light deflections, as observed in the intergalactic gravitational lensing, are also ascribed to the existence of this extra matter. This is as known as the Dark Matter problem. 

The expected Keplerian motions of stellar objects in the galactic spiral arms are something that are predicted by solving the Einstein Field Equation of General Relativity under the visibly observed mass-energy distributions in the spiral galaxies. The deviations from these expectations are attributed to, by most people, the existence of dark matter. But it could well happen that the General Theory of Relativity is not sophisticated enough to explain all these observed phenomena in gravity. Here we are taking this point of view, and are proceeding to see if we can obtain an explanation of these anomalous astronomical phenomena by going beyond Einstein's General Theory of Relativity.

\section{A Yang-Mills type gauge theory of gravity}

There, by now, exist many modifications to Einstein's General Theory of Relativity so as to tackle gravity. Given the great success of the Yang-Mills gauge principle in describing the electroweak and strong interactions, it is compelling to apply the same fundamental symmetry principle to the gravitational interaction. As rigorously derived in our foundational work \cite{ref:5} based on the Erlangen Program and the local $GL(4, R)$ affine symmetry, the Yang-Mills type of action that describes gravitation in the presence of a non-dynamical background world metric is given by:
\begin{equation}
\label{eq:YM_action}
\begin{split}
  {\rm S_{YM}}\left[g, \Gamma, \partial\Gamma \right] = & \int d^4x \sqrt{-g} \frac{1}{2\kappa} \\
  & \times g^{\mu\mu'} g^{\nu\nu'} ( R^{\lambda}_{\ \sigma\mu\nu}R^{\sigma}_{\ \ \lambda\mu'\nu'} ),
\end{split}
\end{equation}
where $g^{\mu\nu}$ is the macroscopic metric, $\kappa$ is the dimensionless coupling constant, and $\Gamma^{\lambda}_{\sigma\nu}$ are the full connections treated as independent dynamical gauge fields. The Riemann curvature tensor here is constructed purely out of the connections by:
\begin{equation}
\label{eq:curvature}
\begin{split}
      R^{\lambda}_{\ \sigma\mu\nu} = & \partial_{\mu}\Gamma^{\lambda}_{\sigma \nu } - \partial_{\nu}\Gamma^{\lambda}_{\sigma \mu } \\
      & + \Gamma^{\lambda}_{\kappa\mu} \Gamma^{\kappa}_{\sigma\nu} - \Gamma^{\lambda}_{\kappa\nu} \Gamma^{\kappa}_{\sigma\mu}.
\end{split}
\end{equation}

This form of the gravitational action has presented itself several times in the history of gravity theories, but carrying profoundly different physical meanings depending on the geometric assumptions. For instance, in Weyl's early proposal \cite{ref:1} and Yang's later formulation \cite{ref:2}, the connections were restricted to be the metric-compatible Christoffel symbols. Such restrictions render the metric the sole dynamical variable, leading to higher-order derivative equations of motion that in\-ev\-i\-ta\-bly suffer from runaway solutions and the Ostrogradski instability \cite{ref:6}. Other approaches, such as Stephenson's \cite{ref:3} or the Poincar$\acute{e}$ Gauge Theory \cite{ref:4}, either artificially suppressed the anti-symmetric parts of the connections (torsion) or restricted the symmetric parts to Christoffel symbols. In our framework, we impose no such a priori geometric restrictions. We regard the full connections (both symmetric and anti-symmetric parts) and the metric as completely independent variables. As demonstrated in Ref. \cite{ref:5}, Eq.~\ref{eq:YM_action} is exactly the Yang-Mills action for the local $GL(4, R)$ gauge group, where the connections are simply the transformed gauge vector potentials. Because the metric acts merely as a background field and the dynamics are governed by the connections through a strictly second-order differential equation, our chosen theory is inherently free from ghosts and Ostrogradski instabilities.

\section{Two legitimate solutions}
The variation of the action given in Eq.~\ref{eq:YM_action} with respect to $g_{\theta\tau}$ will give the so called Stephenson Equation~\cite{ref:3}
\begin{equation}
\label{eq:Step_eq}
\begin{split}
   H_{\theta\tau} \equiv & R_{\  \sigma\theta\rho}^{\lambda} R_{ \ \lambda\tau}^{\sigma \ \ \ \rho} -\frac{1}{4}g_{\theta \tau}R^{\lambda \ \xi\rho}_{\ \sigma }R_{\ \lambda\xi\rho}^{\sigma } \\
   = & \frac{1}{2\kappa} T_{\theta\tau},
\end{split}
\end{equation}
which as we can see, has an algebraic expression for the various components of the Riemann curvature tensor on the left-hand-side of the equation. The variation of the same action with respect to the connections will give the so called Stephenson-Kilmister-Yang equation~\cite{ref:2,ref:3,ref:8}
\begin{equation}
\label{eq:SKY_eq}
   \nabla_{\rho}(\Gamma)(\sqrt{-g}R^{ \ \beta\rho\lambda}_{\sigma }) = \frac{1}{\kappa} \sqrt{-g}S^{\ \beta\lambda}_{\sigma}.
\end{equation}
The $T_{\theta\tau}$ and $S^{\ \beta\lambda}_{\sigma}$ are respectively the metric energy-momentum tensor and the gauge current tensor of the source and the test object, coming from varying the matter part of the action with respect to the metric and the connections. $\nabla_{\rho}(\Gamma)$ here denotes covariant differentiation with the connections $\Gamma$. 
     
If we choose to describe gravity with the above action, then it is mandatory for us to seek for all the possible simultaneous solutions to Eq.~\ref{eq:Step_eq} and Eq.~\ref{eq:SKY_eq}. These are complex equations, but we are fortunate enough to locate two spherically symmetric vacuum ($T_{\theta\tau}$ = 0 and $S^{\ \beta\lambda}_{\sigma}$ = 0) solutions in the literatures. They are both asymptotically Minkowskian and singular at $r$ = 0 (where the source of the gravity is supposed to sit at). In these two solutions, the anti-symmetric parts of the connections are zero (torsion free) and the symmetric parts happen to be the same as the Christoffel symbols calculated from their respective metrics. These metrics are the Schwarz\-schild metric
\begin{equation}
   \label{eq:S_metric}
   ds^2 = \left( 1 - \frac{2GM}{r} \right)dt^2 - \left( 1 - \frac{2GM}{r} \right)^{-1} dr^2 - r^2 d\Omega^2,
\end{equation}
and the Thompson-Pirani-Pavelle-Ni (TPPN) metric~\cite{ref:7,ref:9}
\begin{equation}
   \label{eq:NPPN_metric}
   ds^2 = \left( 1 + \frac{ G'M' }{r} \right)^{-2}dt^2 - \left( 1 + \frac{ G'M' }{r} \right)^{-2} dr^2 - r^2 d\Omega^2,
\end{equation}  
where $GM$ and $G'M'$ are the integration constants for the solutions. In these two solutions, the metrics happen to be compatible with their respective connections. But we have to bear in our mind that we have not assumed metric compatibility, as a priori, in our formulation of the theory. The existence of some solutions which are metric compatible does not mean that the theory is a higher derivative theory.

The TPPN solution was first discovered as a solution to the vacuum Eq.~\ref{eq:SKY_eq}, with symmetric connections and with no reference to Eq.~\ref{eq:Step_eq}. It was later shown by Baekler, Yasskin, Ni, and Fairchild~\cite{ref:9}, and by Hsu and Yeung~\cite{ref:10} that this solution satisfies vacuum Eq.~\ref{eq:SKY_eq} for the full connections. It is also easy to see that these metrics will also satisfy the vacuum Eq.~\ref{eq:Step_eq}~\cite{ref:9,ref:10}. In fact, it was shown~\cite{ref:9,ref:10} that the Schwarz\-schild and the TPPN solutions are the only two possible simultaneous solutions to vacuum Eq.~\ref{eq:Step_eq} and vacuum Eq.~\ref{eq:SKY_eq} for spherical symmetric situation under the compatibility ansatz.

The most straight forward way to show the validity of the solution given in Eq.~\ref{eq:S_metric} and Eq.~\ref{eq:NPPN_metric} is to cast vacuum Eq.~\ref{eq:Step_eq} and vacuum Eq.~\ref{eq:SKY_eq} in their spherical symmetric forms. In the Appendix, we will give vacuum Eq.~\ref{eq:Step_eq} and vacuum Eq.~\ref{eq:SKY_eq} in their spherical symmetric forms, and we can see that both solutions satisfy the equations. We don't know whether there exist solutions to our theory that have anti-symmetric components in the connections, or whether there exist solutions in which the metric is not compatible with the connections. But we shall assume that these yet undiscovered solutions will play no role in the discussions of the following physical phenomena.

\section{Interpretation of the dual vacuum solutions in the Yang-Mills framework}

In Einstein's General Relativity, the vacuum is geometrically trivial in the absence of mass-energy sources. However, in the Yang-Mills type gauge theory of gravity governed by the action in Eq. 1, the full connections $\Gamma^{\lambda}_{\sigma\nu}$ act as independent dynamical gauge fields. Similar to the gluon fields in Quantum Chromodynamics (QCD), these gravitational gauge fields possess non-linear self-interactions. Consequently, the theory admits non-trivial vacuum configurations where the gauge field itself carries an effective energy-momentum, modifying the spacetime geometry at macroscopic scales.

We interpret the existence of the two legitimate spherically symmetric solutions (Eq.~\ref{eq:S_metric} and Eq.~\ref{eq:NPPN_metric}) as the manifestation of this rich gauge structure. The Schwarz\-schild metric ($\bar{g}$) corresponds to the standard geometrical response to localized baryonic mass $M$. In contrast, the TPPN metric ($g'$) represents a non-perturbative vacuum configuration driven by the self-interaction of the gravitational gauge fields. The integration constant $M'$ in the TPPN solution does not represent any physical "dark" particles; rather, it characterizes the total effective energy enclosed within a region due to the gauge field self-interaction, scaled by an effective gauge coupling constant $G'$.

To describe the macroscopic geodesic motion of a standard baryonic test particle (e.g., a regular star of mass $m$) in a galaxy, we must consider the combined geometric effect of both the baryonic source and the gauge field background. As strictly justified by the weak-field asymptotic expansion in Appendix A, the effective spacetime geometry $g_{\mathrm{eff}}$ experienced by the baryonic matter is a linear superposition of the two metric states:
\begin{equation}
\label{eq:superposition}
g_{\mathrm{eff}} = (1 - \alpha)\bar{g} + \alpha g'
\end{equation}
where $\alpha \ll 1$ is a universal, dimensionless coupling parameter characterizing the relative strength of the non-linear Yang-Mills gauge background to the standard Newtonian background in the galactic halo.

Because the test star consists solely of regular baryonic matter, its motion is entirely determined by the geodesic equation of this effective metric $g_{\mathrm{eff}}$. In the low-velocity limit, the effective radial acceleration is given by:
\begin{equation}
\frac{d^2r}{dt^2} = \frac{1}{2} g_{\mathrm{eff}}^{rr} \frac{\partial (g_{\mathrm{eff}})_{00}}{\partial r}
\end{equation}
Substituting the superposed metric into this equation, the centrifugal force required to maintain a circular orbit of radius $r$ for a star of mass $m$ is exactly balanced by the modified gravitational pull:
\begin{equation}
\label{eq:v_squ}
\begin{split}
m \frac{v^2}{r} = & m \bigg[ (1-\alpha) \frac{GM}{r^2} \\
& + \alpha \frac{G'M'}{r^2} \left(1+\frac{G'M'}{r}\right)^{-1} \bigg] 
\end{split}
\end{equation}
which immediately yields the modified rotation speed $v$ as a function of $r$:
\begin{equation}
v^2 = (1-\alpha)\frac{GM}{r} + \alpha \frac{G'M'}{r+G'M'}
\end{equation}
This formulation demonstrates that a single baryonic star, without binding to any exotic dark matter particles, will naturally exhibit non-Keplerian kinematics due to the intrinsic gauge field self-interactions of the Yang-Mills gravitational vacuum.

\section{Predictions of the Universal rotation curves for spiral galaxies}

The modified kinematic relation derived above can be immediately applied to describe the rotation curves of spiral galaxies. In our modified gravity framework, the galactic disk is composed solely of regular baryonic matter of mass $M$. The galactic halo is no longer hypothesized as a cloud of exotic dark matter particles; rather, it is described as a macroscopic region permeated by a non-trivial Yang-Mills gauge field configuration, which manifests as a uniform effective energy density $\rho_{\mathrm{YM}}$ at the background level. 

The effective gauge mass enclosed within a radius $r$ is given by $M_{\mathrm{YM}}(r) = \frac{4\pi}{3}r^3\rho_{\mathrm{YM}}$. The spiral arm structure of the baryonic matter contributes in the form of modified Bessel functions. Thus, the total rotation speed squared $v^2$ under the effective metric $g_{\mathrm{eff}}$ is the weighted sum of the Newtonian disk contribution and the Yang-Mills geometric contribution:
\begin{eqnarray}
\label{eq:v_squ_new}
v^2 &=& (1-\alpha)v_{\mathrm{d}}^2 \nonumber \\
&& + \alpha \frac{G' \frac{4\pi}{3}\rho_{\mathrm{YM}}r^3}{r + G'\frac{4\pi}{3}\rho_{\mathrm{YM}}r^3} \nonumber \\
&=& (1-\alpha) \Big[ 4\pi G\Sigma_0 h y^2 \big[ I_0(y)K_0(y) \nonumber \\
&& - I_1(y)K_1(y) \big] \Big] + \alpha \frac{G^* r^3}{r + G^* r^3}
\end{eqnarray}
where $I_i, K_i$ are modified Bessel functions of the first and second kinds, and $y = \frac{r}{2h}$. The asymptotic gauge coupling weight $\alpha$ determines the mixing of the two metric components, and $G^* = G'\frac{4\pi}{3}\rho_{\mathrm{YM}}$ represents the effective coupling strength of the gauge field background.

\begin{figure*}[htbp]
\centering
\begin{tabular}{cc}
\includegraphics[width=0.48\textwidth]{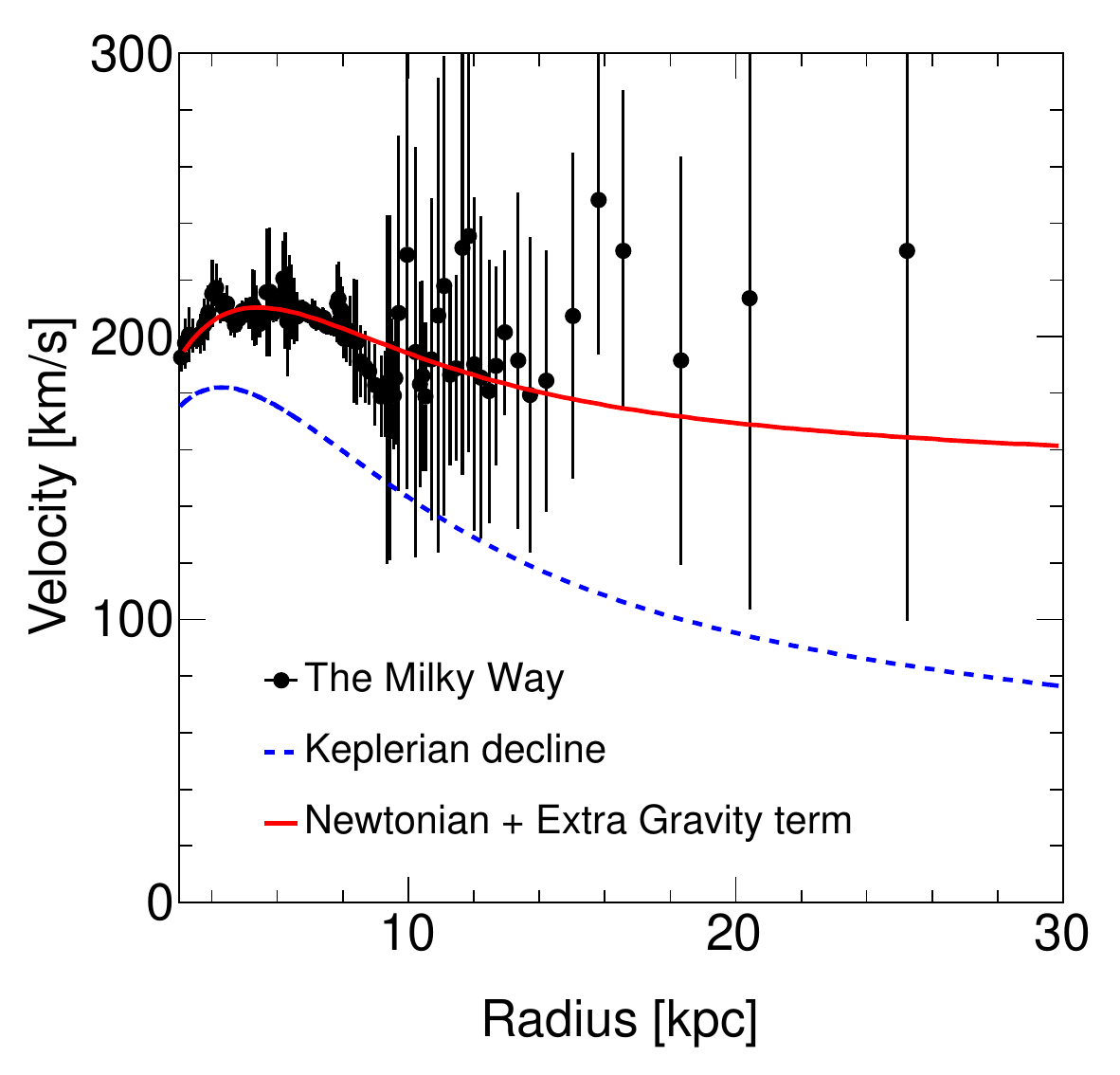} &
\includegraphics[width=0.48\textwidth]{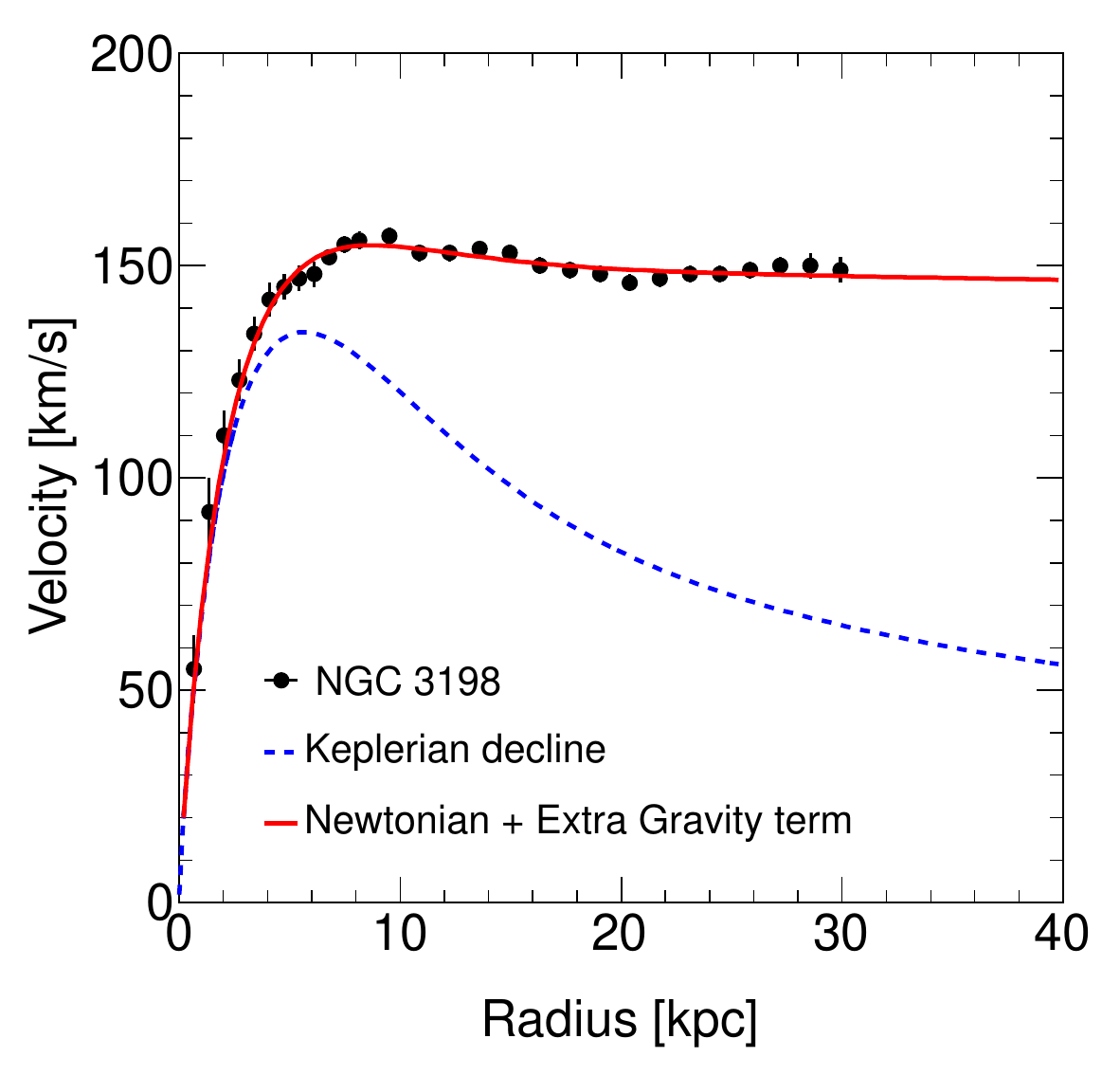} \\
(a) & (b) \\
\includegraphics[width=0.48\textwidth]{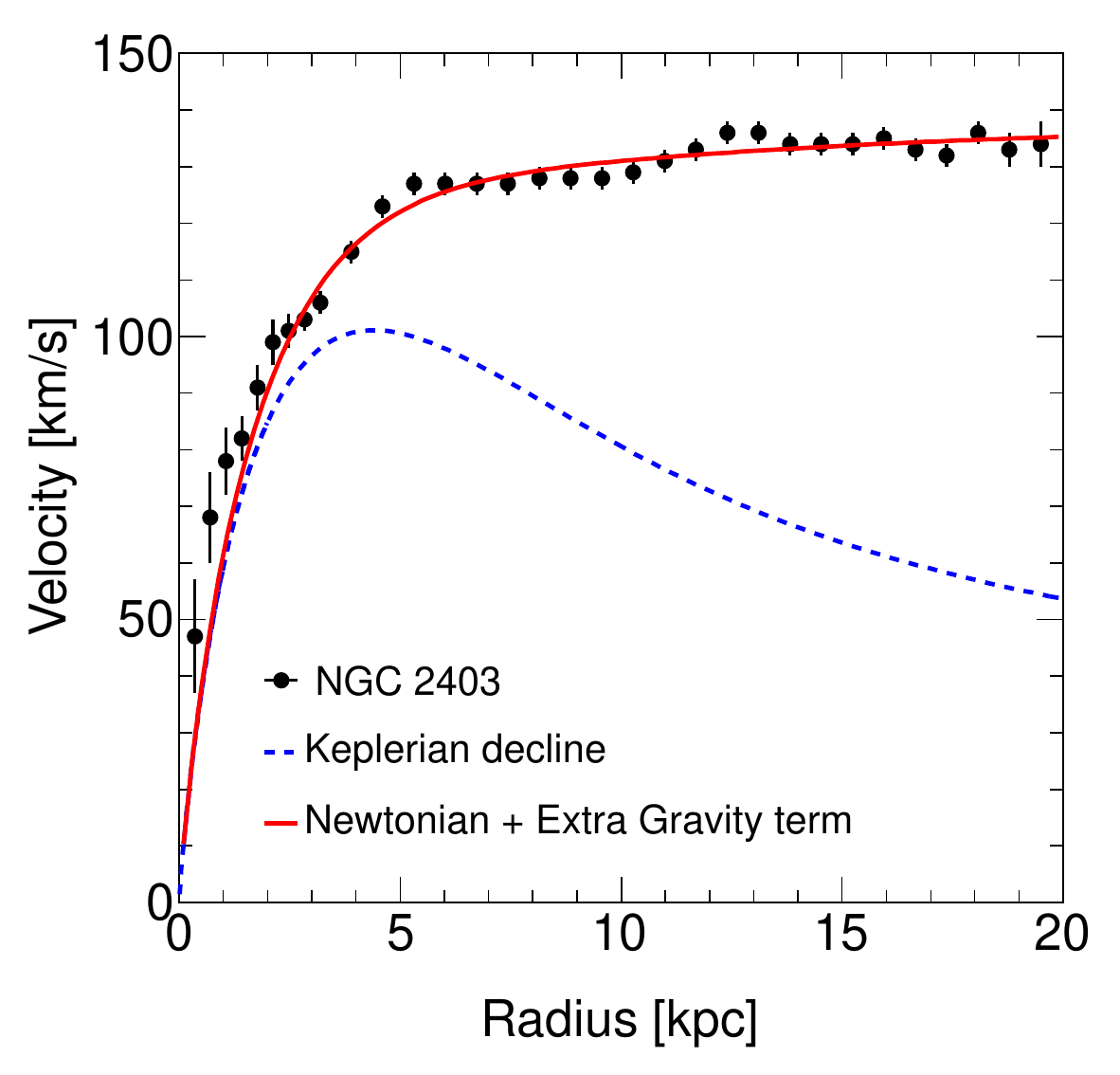} &
\includegraphics[width=0.48\textwidth]{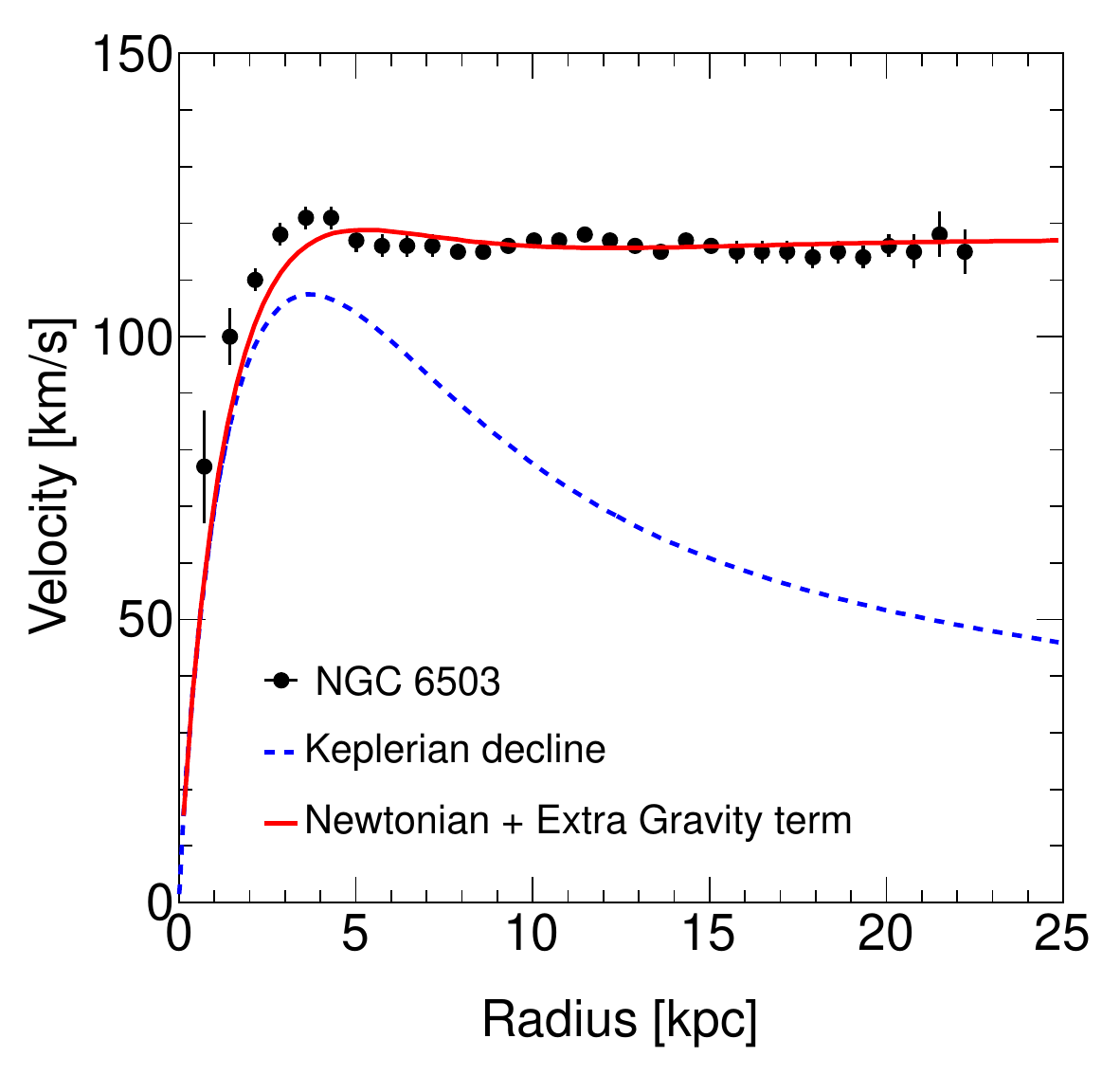} \\
(c) & (d)
\end{tabular}
\caption{The predicted relationship between the galactic rotation speed $v$ and the distance $r$ from a combined influence of the Newtonian force and the new geometric gauge field background: (a) The Milky Way, (b) NGC 3198, (c) NGC 2403 and (d) NGC 6503.}
\label{fig:1}
\end{figure*}

This expression structurally matches the empirical formula proposed by Salucci et al.~\cite{ref:13}, who analyzed a large sample of spiral galaxies. By fitting well-known galaxies~\cite{ref:14,ref:15,ref:16,ref:17,ref:18} (The Milky Way, NGC 3198, NGC 2403, and NGC 6503), we extract a universal effective coupling $G^* \approx 10^{-2}\text{ kpc}^{-2}$ and a universal gauge coupling weight $\alpha \approx 2 \times 10^{-9}$ for radii ranging from 3 kpc to 30 kpc. The results are shown in Fig.~\ref{fig:1} and Table~\ref{table:1}.

\begin{table*}[htbp]
\centering
\begin{tabular}{lcccc}
\hline\noalign{\smallskip}
 & The Milky Way & NGC 3198 & NGC 2403 & NGC 6503 \\
\noalign{\smallskip}\hline\noalign{\smallskip}
$G\Sigma_{0}\ [km^{2}s^{-2}kpc^{-1}]$ & $6.8\times10^{3}$ & $2.8\times10^{3}$ & $2.1\times10^{3}$ & $2.8\times10^{3}$ \\
$h$ [kpc] & 2.0 & 2.63 & 2.05 & 1.72 \\
$G^{*}\ [kpc^{-2}]$ & $5.0\times10^{-2}$ & $9.2\times10^{-3}$ & $1.4\times10^{-2}$ & $1.3\times10^{-2}$ \\
$\alpha$ & $2.3\times10^{-9}$ & $2.2\times10^{-9}$ & $2.0\times10^{-9}$ & $1.4\times10^{-9}$ \\
\noalign{\smallskip}\hline
\end{tabular}
\caption{The fitting parameters for some galaxies. Note that the gauge coupling weight $\alpha$ structurally replaces the conventional dark matter mass ratio.}
\label{table:1}
\end{table*}

\begin{figure}[htbp]
\centering
\includegraphics[width=0.48\textwidth]{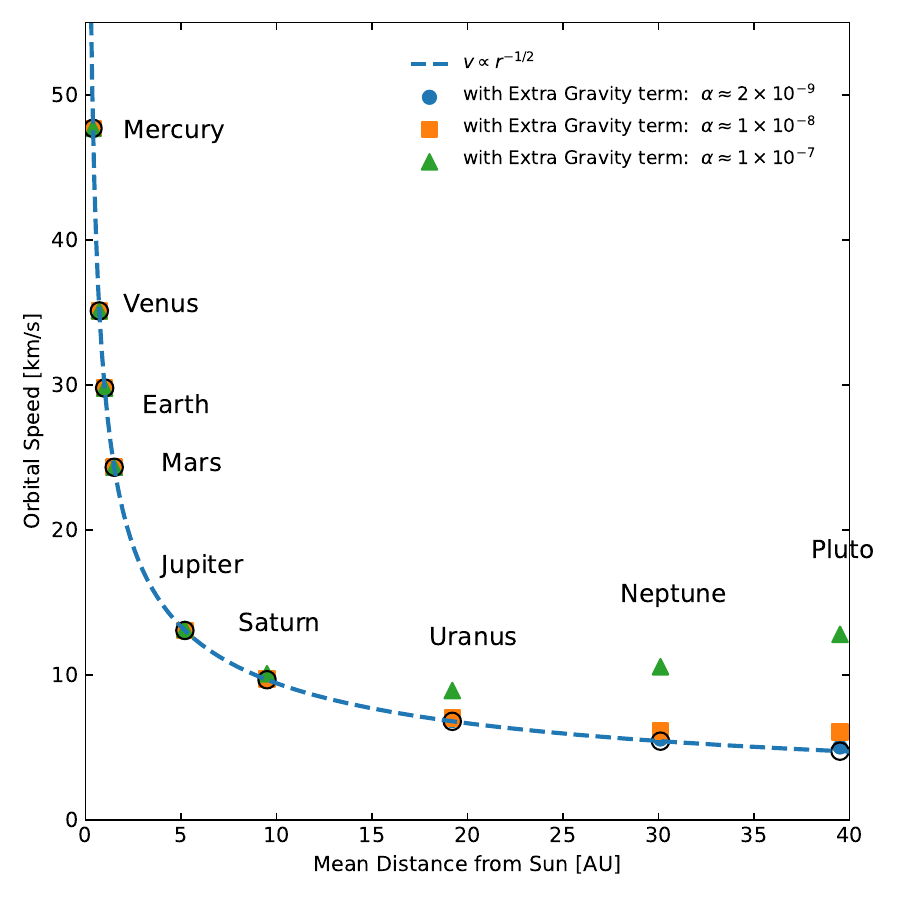}
\caption{The orbital speeds of the planets in the solar system predicted with different effective gauge coupling weights $\alpha$. The geometric modifications are completely suppressed at this scale.}
\label{fig:2}
\end{figure}

It is crucial to verify the influence of this effective gauge background on solar system dynamics. At the scale of planetary orbits, the enclosed effective gauge energy $M_{\mathrm{YM}}$ is vanishingly small. Using the universal values of $G^*$ and $\alpha$ obtained from galactic fits, the predicted planetary rotation speeds remain precisely consistent with standard Keplerian kinematics. Figure~\ref{fig:2} demonstrates that the non-linear gauge modifications are safely suppressed at the solar system scale, leaving the fundamental tests of General Relativity intact.

\section{Right amounts of intergalactic gravitational lensing}
Next, we address the phenomenon of anomalous light deflection observed in intergalactic gravitational lensing, conventionally attributed to dark matter. We take the galaxy cluster Abell 1689 as an illustration. In our framework, the cluster halo is not a collection of invisible particles but a large-scale non-trivial Yang-Mills gauge field configuration overlapping with the cluster's extent (approximately $R = 300\text{ kpc}$).

The azimuthal angle swept by light traveling from the edge of the halo $R$ to the point of closest approach $r_0$, under the influence of the TPPN geometric component of our effective metric, is given by~\cite{ref:19}:
\begin{eqnarray}
\Delta\varphi &=& \int_{r_0}^{R} \frac{(r_0 + G'M_{\mathrm{YM}})}{(r + G'M_{\mathrm{YM}})} \nonumber \\
&& \times \frac{dr}{\sqrt{(r + G'M_{\mathrm{YM}})^2 - (r_0 + G'M_{\mathrm{YM}})^2}} \nonumber \\
\end{eqnarray}
which yields:
\begin{equation}
\Delta\varphi = \sec^{-1}\frac{1+\beta R}{1+\beta r_0}
\end{equation}
where $\beta = (G'M_{\mathrm{YM}})^{-1}$.

For a localized point source, as $R \rightarrow \infty$, the deflection approaches standard predictions. However, for an extended gauge field energy distribution within the halo, the light ray penetrates regions of varying effective mass $M_{\mathrm{YM}}(r)$. The variation in the azimuthal angle due to this extended geometric configuration is:
\begin{eqnarray}
\delta(\Delta\varphi) &=& \frac{\partial(\Delta\varphi)}{\partial M_{\mathrm{YM}}} \delta M_{\mathrm{YM}} \nonumber \\
&=& \frac{3}{\sqrt{2}}(R-r_0)^{\frac{3}{2}} (G^*)^{-\frac{1}{2}} R^{-\frac{5}{2}}
\end{eqnarray}
Using the physical parameters for Abell 1689 ($R = 300\text{ kpc}$, $R-r_0 = 30\text{ kpc}$) and the universal gauge coupling $G^* = 10^{-2}\text{ kpc}^{-2}$ derived independently from galactic rotation curves, we obtain a deflection angle of $4 \times 10^{-3}$ rad. This value quantitatively accounts for the excess lensing traditionally attributed to dark matter, achieving consistency across different astronomical observables without introducing any new particles.

\section{Conclusion}

The Yang-Mills type gauge theory of gravity presents a richer geometric structure than Einstein's General Theory of Relativity by elevating the full connections to dynamical gauge fields. We have shown that the non-linear self-interactions of these gravitational gauge fields give rise to non-trivial vacuum solutions, independent of any hypothetical dark sector particles.

From a phenomenological perspective, by taking the effective spacetime geometry as a linear superposition of the standard baryonic response (Schwarz\-schild) and the gauge background response (TPPN) in the weak-field limit, our semi-empirical model successfully reproduces the universal galactic rotation curves and accounts for the anomalous intergalactic gravitational lensing of \linebreak Abell 1689. This is achieved through a universal effective \linebreak gauge coupling $G^* \approx 10^{-2}\text{ kpc}^{-2}$ and a macroscopic geometric mixing weight $\alpha \approx 2 \times 10^{-9}$. 

While the exact micro-physical derivation of the gauge parameters ($\alpha$ and $G^*$) via renormalization group flow requires further theoretical development, the current phenomenological success is highly non-trivial. These quantitative results strongly suggest that the phenomena currently ascribed to "Dark Matter" may simply be macroscopic manifestations of the intrinsic non-linear geometric properties and vacuum configurations of the Yang-Mills gauge theory of gravity. The implications of this geometric framework on cosmological scales and the Dark Energy problem are addressed in a separate work.

\appendix
\section{Asymptotic Justification for Metric Superposition in the Weak-Field Limit}
In this appendix we will cast the Stephenson Equation and the Stephenson-Kilmister-Yang Equation in static spherical symmetric forms. The metric is taken to be the form:
\begin{equation}
  \label{eq:gen_metric_2}  
  ds^2 = B(r)dt^2 - A(r)dr^2 -r^2d\Omega^2.
\end{equation}
There are three independent, non-vanishing components for the left-hand-side of Eq.~\ref{eq:Step_eq}. They are, the $tt$ component, 
\begin{eqnarray}
    \label{eq:tt_component}
   && 8r^2A^2B^2B'^2 + r^4(A'B'B - AB'^2 \nonumber  \\
   && - 2AB'' + 2AB'^2)^2 \nonumber  \\
   && - 16A^2B^4(A-1)^2 - 8r^2A'^2B^4 = 0, 
\end{eqnarray}
the $rr$ component:
\begin{eqnarray}
    \label{eq:rr_component}
   && 8r^2A'^2B^4 + r^4(A'B'B - AB'^2 \nonumber  \\
   && - 2ABB'' + 2AB'^2)^2  \nonumber  \\
   && - 16A^2B^4(A-1)^2 - 8r^2A^2B^2B'^2 = 0,
\end{eqnarray}
and the $\theta\theta$ component (same as the $\phi\phi$ component),
\begin{eqnarray}
    \label{eq:thetatheta_component}
    && r^4(A'B'B - AB'^2 - 2ABB'' + 2AB'^2)^2  \nonumber  \\
    && - 16A^2B^4(A-1)^2 = 0.
\end{eqnarray}
There are two independent components for the left-hand-side of Eq.~\ref{eq:SKY_eq}. They are~\cite{ref:7}
\begin{eqnarray}
    \label{eq:SKY_ind_1}
    &&r^2(2A^2B^2B^{\prime\prime\prime} - 4A^2B''B'B - 3 AA'B^2B'' \nonumber  \\ 
    &&    + 2A^2B'^3 + 2AA'BB'^2 - AA''B^2B' \nonumber \\
    &&    + 2A'^2B^2B' ) + 2rAB(2ABB'' - AB'^2 \nonumber \\
    &&    - A'BB') - 4A^2B^2B' = 0,
\end{eqnarray}
\begin{eqnarray}
    \label{eq:SKY_ind_2}
    && r^2(2AA''B^2 + AA'BB' - 4A'^2B^2) \nonumber \\
    && - 4A^2B^2(A-1) =  0.
\end{eqnarray}
Subtracting Eq.~\ref{eq:rr_component} from Eq.~\ref{eq:tt_component} will give 
\begin{eqnarray}
    \label{eq:subtract_eq}
    && 8r^2A^2B^2B'^2 - 8r^2A'^{2}B^4 \nonumber \\
    && = 8r^2A'^2B^4 - 8r^2A^2B^2B'^2
\end{eqnarray}
and will lead to 
\begin{eqnarray}
    \label{eq:subtract_eq_2}
    \frac{B'}{B} = \pm \frac{A'}{A}.
\end{eqnarray}
Eq.~\ref{eq:subtract_eq_2} are the two solutions for the Schwarz\-schild metric and the TPPN metric, respectively.

While the linear superposition of metrics (Eq.~\ref{eq:superposition}) is not an exact solution to these highly non-linear generalized Stephenson equations, its phenomenological validity for galactic dynamics must be evaluated by quantifying the residual cross-terms in the far-field region (galactic halo scale). Let us define the weight factors $w_1 = (1-\alpha)$ and $w_2 = \alpha$. At macroscopic distances $r \gg GM$ and $r \gg G'M'$, we can expand the exact solutions $\bar{B}(r)$, $\bar{A}(r)$ (Schwarz\-schild) and $B'(r)$, $A'(r)$ (TPPN) in powers of $1/r$:
\begin{eqnarray}
\bar{B}(r) &=& 1 - \frac{2GM}{r} + \mathcal{O}(r^{-2}), \\
\bar{A}(r) &=& 1 + \frac{2GM}{r} + \mathcal{O}(r^{-2}) \\
B'(r) &=& 1 - \frac{2G'M'}{r} + \mathcal{O}(r^{-2}), \\
A'(r) &=& 1 - \frac{2G'M'}{r} + \mathcal{O}(r^{-2})
\end{eqnarray}

Substituting these expansions into the superposed metric $B(r) = w_1 \bar{B} + w_2 B'$ and $A(r) = w_1 \bar{A} + w_2 A'$ yields:
\begin{eqnarray}
B(r) &=& 1 - \frac{2}{r}(w_1 GM + w_2 G'M') + \mathcal{O}(r^{-2}) \\
A(r) &=& 1 + \frac{2}{r}(w_1 GM - w_2 G'M') + \mathcal{O}(r^{-2})
\end{eqnarray}

We now insert this superposed Ansatz back into the exact non-linear operator $H_{tt}[g]$ defined by the left-hand side of Eq.~\ref{eq:Step_eq}. The exact solutions identically satisfy $H_{tt}[\bar{g}] = 0$ and $H_{tt}[g'] = 0$. The evaluation of $H_{tt}[g]$ thus isolates the non-linear cross-terms, which constitute the residual error $E_{tt}(r)$ of our approximation. The dominant non-linear terms depend on quantities such as $(A-1)^2$, $(B')^2$, and $A'$. In the weak-field limit, the spatial derivatives scale as $\partial_r \sim \mathcal{O}(r^{-2})$. Consequently, the cross-terms taking the form of $\alpha(A_S-1)A_T$ or similar products will scale strictly as:
\begin{equation}
E_{tt}(r) \sim \mathcal{O}\left( \frac{(GM)(G'M')}{r^4} \right)
\end{equation}

At typical galactic halo scales ($r \sim 10\text{ kpc}$), the dimensionless gravitational potential $\frac{GM}{r}$ is of the order of $10^{-6}$. The residual non-linear terms governing the error of the metric superposition are therefore suppressed by a factor of at least $10^{-12}$ compared to the leading kinematic terms. This rigorous asymptotic suppression guarantees that the mixed source configuration introduces negligible theoretical error when fitting the macroscopic galactic rotation curves.

\section*{Acknowledgements}
The authors acknowledge support from Academia Sinica and National Cheng Kung University (NCKU). We would like to thank Professor F. W. Hehl, Professor James Nester, Professor T. C. Yuan and Professor Daniel Wilkins for valuable discussions.

\end{document}